\documentclass[copyright,creativecommons,noderivs,noncommercial]{eptcs}

\usepackage[figuresright]{rotating}
\usepackage[latin1]{inputenc}
\usepackage{alltt,calc}
\usepackage{textcomp}
\usepackage{ifthen}
\usepackage{listings}
\usepackage{pxfonts}
\usepackage{txfonts}
\usepackage{amssymb}
\usepackage{breakurl}

\newcommand{\codeIsEqualToInContext}[2]{\quad%
     \ensuremath{=_{#1,#2}}\quad\hspace{-0.6em}}

\newcommand{\subclassOp}{\ensuremath{\leq} }
\newcommand{\notSubclassOp}{\ensuremath{\nleq} }

\newcommand{\cc}[1]{\ensuremath{#1}}
\newcommand{\keywordFont}[1]{\ensuremath{\mathbf{#1}}}

\newcommand{\nullable}{\keywordFont{nullable}}
\newcommand{\also}{\keywordFont{also}}
\newcommand{\invariant}{\keywordFont{invariant}}

\newcommand{\requires}{\keywordFont{requires}}

\newcommand{\ensures}{\keywordFont{ensures}}

\newcommand{\class}{\keywordFont{class}}

\newcommand{\extends}{\keywordFont{extends}}

\newcommand{\classEnd}{\keywordFont{\}}}

\newcommand{\If}{\keywordFont{if}}
\newcommand{\Object}{\keywordFont{Object}}

\newcommand{\This}{\keywordFont{this}}
\newcommand{\Super}{\keywordFont{super}}
\newcommand{\instanceof}{\keywordFont{instanceof}}

\newcommand{\Else}{\keywordFont{else}}

\newcommand{\providedJava}{\item[\textbf{\textit{Java:}}]}
\newcommand{\providedJml}{\item[\textbf{\textit{JML:}}]}

\newcommand{\specStart}{$/*@$}

\newcommand{\specEnd}{$@*/$}

\newcommand{\subescrito}[2]{$#1_#2$}

\newcommand{\refJmlLaw}[1]{\textbf{Law #1}}

\newlength{\codeLength}
\newsavebox{\fminibox}
\newlength{\fminilength}
\newenvironment{fminipage}[1][\linewidth]
   {\setlength{\fminilength}{#1-2\fboxsep-2\fboxrule}
    \begin{lrbox}{\fminibox}\begin{minipage}{\fminilength}}
   {\end{minipage}\end{lrbox}\noindent\fbox{\usebox{\fminibox}}}
\newenvironment{framedCode}
     {%
      \setlength{\codeLength}{0.41\linewidth}
      \begin{fminipage}[\codeLength]      
      \begin{alltt}}
     {\end{alltt}
      \end{fminipage}}

\newtheorem{law}{Law}

\newtheorem{jmllaw}{Law}

\newtheorem{javajmllaw}{Law}

\newcommand{\lawName}[1]{\ensuremath{\langle}\textit{#1}%
                         \ensuremath{\rangle}}

\newcommand{\lawJmlName}[1]{\hspace{-0.5em} \ensuremath{\langle}\textit{#1}%
                         \ensuremath{\rangle}}

\newenvironment{LawJml}[1]
{\begin{jmllaw} \noindent \normalfont \lawJmlName{#1} \\ \noindent }
{\hfill $\Box$\end{jmllaw}}

\newcommand{\lawJavaJmlName}[1]{\hspace{-0.5em}\ensuremath{\langle}\textit{#1}%
                         \ensuremath{\rangle}}

\lstdefinelanguage[JML]{Java}[]{Java}%
       {
        comment=[l]{//\ },
        morecomment=[s]{/*\ }{*/},        
        morecomment=[s]{/**}{*/},
        classoffset=1,
        morekeywords={abrupt_behavior,abrupt_behaviour,
         accessible,accessible_redundantly,also,assert,assert_redundantly,
         assignable,assignable_redundantly,assume,assume_redundantly,
         axiom,behavior,behaviour,breaks,breaks_redundantly,
         callable,callable_redundantly,captures,captures_redundantly,
         choose,choose_if,code,code_bigint_math,code_java_math,
         code_safe_math,constraint,constraint_redundantly,constructor,
         continues,continues_redundantly,decreases,decreases_redundantly,
         decreasing,decreasing_redundantly,diverges,diverges_redundantly,
         duration,duration_redundantly,ensures,ensures_redundantly,
         example,exceptional_behavior,exceptional_behaviour,
         exceptional_example,exsures,exsures_redundantly,extract,field,
         forall,for_example,ghost,helper,hence_by,hence_by_redundantly,
         implies_that,in,in_redundantly,initializer,initially,instance,
         invariant,invariant_redundantly,loop_invariant,
         loop_invariant_redundantly,maintaining,maintaining_redundantly,
         maps,maps_redundantly,measured_by,measured_by_redundantly,method,
         model,model_program,modifiable,modifiable_redundantly,modifies,
         modifies_redundantly,monitored,monitors_for,non_null,
         normal_behavior,normal_behaviour,normal_example,nowarn,
         nullable,nullable_by_default,old,or,post,post_redundantly,
         pre,pre_redundantly,pure,readable,refine,refines,refining,represents,
         represents_redundantly,requires,requires_redundantly,
         returns,returns_redundantly,set,signals,signals_only,
         signals_only_redundantly,signals_redundantly,spec_bigint_math,
         spec_java_math,spec_protected,spec_public,spec_safe_math,
         static_initializer,uninitialized,unreachable,weakly,
         when,when_redundantly,working_space,working_space_redundantly,
         writable,result
        },
        morekeywords={rep,peer,readonly},
        keywordsprefix=\\,
        otherkeywords={<:,<-,->,..,<==,==>,<==>,<=!=>},
        classoffset=0 
}

\newtheorem{mydefinition}{Definition}

\title{Object-oriented Programming Laws for Annotated Java Programs}

\author{Gabriel Falconieri Freitas 
\institute{Universidade de Pernambuco (UPE), Brasil}
\email{grff@dsc.upe.br}
\and
M\'{a}rcio Corn\'{e}lio
\institute{Universidade de Pernambuco (UPE), Brasil}
\institute{Universidade Federal Rural de Pernambuco (UFRPE), Brasil}
\email{marciocornelio@acm.org}
\and
Tiago Massoni \qquad Rohit Gheyi
\institute{Universidade Federal de Campina Grande (UFCG), Brasil}
\email{\{massoni, rohit\}@dsc.ufcg.edu.br}
}

\def\titlerunning{Object-oriented Programming Laws for Annotated Java Programs}
\def\authorrunning{Gabriel Falconieri Freitas et al.}

\begin{document}
\maketitle

\sloppy

\begin{abstract}
Object-oriented programming laws have been proposed in the context of languages
that are not combined with a behavioral interface specification language (BISL).
The strong dependence between source-code and interface specifications may cause a number of difficulties when
transforming programs. 
In this paper we introduce a set of programming laws for  object-oriented languages like Java combined
with the Java Modeling Language (JML). The set of laws deals with object-oriented features taking into account their specifications.
Some laws deal only with features of the specification language. These laws constitute a
set of small transformations for the development of more elaborate ones like refactorings.
\end{abstract}

\section{Introduction}

Software changes constantly due to maintenance that leads to correction
of fails or just to improve functionalities. However, some changes can
take place to achieve quality factors like reuse and legibility. In
these cases, changes should not alter the software behavior but only its
internal structure. Improving the internal software structure is an
activity known as refactoring~\cite{Fowler}. To avoid errors due to
modifications, every change has to be done following a discipline.

This discipline can be achieved by programming laws,  as guidelines to informal programming practices, 
establishing a basis for formal and rigorous program development.  They
are largely known for imperative programming~\cite{hoare,Morgan:book}.
Also, functional programming and logic programming have a set of laws
described by Bird and de Moor~\cite{BM97}  and
Seres~\cite{Seres:thesis}, respectively. Laws of object-oriented
programming have also been addressed in~\cite{Borba04,Duarte08,CCS09}.

Design by Contract (DbC)~\cite{Meyer92applyingdesign} is a development
methodology that aims at the construction of reliable object-oriented
systems. Its basic idea is that a contract is established among classes
of a system.  In this way, software developers should formally specify
what is required and ensured by methods and types. The Java Modeling
Language (JML)~\cite{JmlRefMan, LC05} is a notation for formally
specifying the behavior of Java classes and methods.

The set of programming laws for object-oriented programming we have
nowadays is designed for program transformation with no relation to
specifications languages designed for DbC. Changes in specification
usually should discharge code updates, maintaining the conformance
between code and specification. On the other hand, changes in program
code may require changes in specifications as the behavior implemented
by code may diverge from the meaning of the original specification. For
instance, moving a redefined method to its superclass can be illegal if
this transformation causes weakening of pre-conditions and strengthening
of post-conditions. Transformation of object-oriented programs with
formal contracts has already been addressed in a rather informal way~\cite{Goldstein06}.

A catalogue of laws (primitive transformations) to deal with 
Java programs annotated with JML has been proposed
in~\cite{Freitas:MSc}, which specifies about 80 laws. Here we present
one law that deals only with JML specifications and two  JML-aware
Java laws that deals with attributes and methods, respectively. A law
that only deals with JML can impose conditions only on JML elements
present in the program, whereas JML-aware Java laws
involve both JML and Java elements for stating conditions.  The three laws
we present here are  catalogued in~\cite{Freitas:MSc}.

In this paper, we  define laws~(Section~\ref{section:laws}) of
object-oriented programming for Java that are aware of specifications
written in JML, which we describe in Section~\ref{section:jml}. 
The laws we present here and other ones present in a more comprehensive
catalogue~\cite{Freitas:MSc} were applied to refactoring a JML-specified version of a core module
of a Manufacturing Execution System (MES)~\cite{MES}.
In Section~\ref{section:soundness}, we present proof of soundness regarding the
JML parts of two laws. We present an example of program transformation
by means of laws in Section~\ref{section:application}. Final remarks appear
in Section~\ref{section:conclusion}.

\section{The Java Modeling Language}
\label{section:jml}

The Java Modeling Language (JML) is a behavioral interface specification
language (BISL)~\cite{JmlRefMan, LC05} tailored to Java~\cite{Java}.
Thus, JML serves to describe names and static information that appear in
Java declarations and how they act, how they behave. JML specifications
are written in the form of \textit{special annotation} comments that are
inserted directly in source code of programs. These comments must begin
with an at-sign (\textbf{@}) and can be written in two ways: by using
\textbf{//@ ...} or \textbf{/*@ ... @*/}.

\begin{figure}[t]
\begin{lstlisting}[frame=single]
	public class PositiveIntegerData {
		//@ private invariant value.intValue() > -1;  (*@\label{jml_example_positive_integer:inv}@*)
		private Integer value;						
		public PositiveIntegerData() { value = new Integer(0); }			

		/*@ requires newValue != null && newValue.intValue() > -1; (*@\label{jml_example_positive_integer:req_registerValue}@*)
		  @ ensures getValue().intValue() == newValue.intValue(); (*@\label{jml_example_positive_integer:ens_registerValue}@*) @*/
		public void registerValue(Integer newValue) { /* ... */ }

		//@ ensures \result != null; (*@\label{jml_example_positive_integer:ens_getValue}@*)		  
		public /*@ pure @*/ Integer getValue() { /* ... */	} (*@\label{jml_example_positive_integer:pure_getValue}@*)		

		/*@ requires getValue() != null; (*@\label{jml_example_positive_integer:req_format}@*)
		  @ ensures !(\result).equals(""); @*/	(*@\label{jml_example_positive_integer:ens_format}@*)
		public String format() { /* ... */ }
	}
\end{lstlisting}
\caption{\label{class:PositiveInteger}Class \lstinline!PositiveIntegerData!}
\end{figure}

In Figure~\ref{class:PositiveInteger}, we present the class
\lstinline!PositiveIntegerData!  that represents positive integers.
We introduce  an instance invariant (Line~\ref{jml_example_positive_integer:inv}), 
which is a predicate that is true in all visible
states of objects of a class~\cite{JmlRefMan}. The invariant in the
example has private visibility and establishes that the 
attribute \lstinline@value@ must always be greater than -1.

JML uses the  \lstinline!requires! clause to specify the obligations of
the caller of a method, what must be true to call a method. For
instance, the precondition of the method  \lstinline!registerValue! requires
the value of the Integer object to be registered  to be greater than -1.

A postcondition specifies
the implementor's obligation, what must be true at the end of a method,
just before it returns to the caller. In JML, the  \lstinline!ensures!
clause introduces a postcondition. For instance, the
Line~\ref{jml_example_positive_integer:ens_registerValue} introduces a normal postcondition that
asserts that the final value of the Integer object we register is the same as
the one the method receives as argument.
The JML modifier \lstinline!pure! (Line~\ref{jml_example_positive_integer:pure_getValue}) indicates
that the method doesn't have any side effects and hence  can appear in specifications. 
In JML, the keyword \lstinline@also@ indicates that a method
is extending the specification it inherits from its supertype.

\section{Laws}
\label{section:laws}


\begin{figure}[t]
\begin{LawJml}{move invariant to superclass}
\label{law:move_invariant_superclass}%
\begin{framedCode} 
\cc{\class B \extends A \{}
  \cc{//@ \invariant} \subescrito{\psi}{1}\cc{;}
  \cc{ads cnts mds}     
\cc{\classEnd}
\cc{\class C \extends B \{}
  \cc{//@ \invariant} \subescrito{\psi'}{1} && \subescrito{\psi}{2}\cc{;}
  \cc{cnts' mds'} 
\cc{\classEnd}
\end{framedCode}%
	\codeIsEqualToInContext{cds}{Main}
\begin{framedCode} 
\cc{\class B \extends A \{}
  \cc{//@ \invariant} \subescrito{\psi}{1} && \subescrito{\psi}{2}\cc{;} 
  \cc{ads cnts mds}    
\cc{\classEnd}
\cc{\class C \extends B \{}
  \cc{//@ \invariant} \subescrito{\psi'}{1}\cc{;}  
  \cc{cnts' mds'}  
\cc{\classEnd}
\end{framedCode}

\begin{description}
\item[\textbf{where}]
       \item[] $\psi_2\ \equiv\ \This\ \instanceof\ C\ ==>\ \psi_\texttt{inv}$
\end{description}

\begin{description}
\item[\textbf{provided}]
\item [($\leftrightarrow$)] $\Super$ does not appear in $\psi_2$.
\item [($\rightarrow$)] $\psi_2$ does not contain occurrences of model fields declared in $C$, nor uncast occurrences of $\This$. 
\end{description}

\end{LawJml}
\end{figure}


Our laws extend object-oriented programming laws from other
works~\cite{Borba04,CCS09,Cornelio04,Duarte08,DBLP:conf/fase/MassoniGB08}.
The laws are written in an equational style. Each side of the equation
corresponds to a template of a well-formed program. Programming laws, in
which left-hand and right-hand sides are related by equality, are a
concise presentation of a pair of laws. These laws precisely indicate
the modifications that can be done to a program, stating their
corresponding side-conditions. In fact, to apply a law, it is
necessary to check (syntactic or semantic) side-conditions to ensure
that the transformation is behavior-preserving and also maintains its
well-formedness. We consider that we are dealing with
only one package and working in a limited open system~\cite{Duarte08},
in which classes of our system can depend on external libraries.

In Java and JML context, we need to guarantee that
source-code continues meeting its specifications written in JML, taking
into account the semantics of JML specifications along with the notion
of specification inheritance~\cite{DBLP:conf/icfem/Leavens06}. 
Here we present a law for invariants written in JML.

A JML-annotated Java program has the format $cds\ Main$, where $cds$ is the set of all 
classes of the program and $Main$ corresponds to the unique 
class in the  program which has a static main method.  
We use $cnts$, $ads$ and $mds$  inside a class to represents the class 
constructors, attributes and methods, respectively. We have to emphasize that  they
may contain the  specifications of each constructor or method. It is 
not necessarily  Java code only, we can also have the corresponding JML specifications.

In the laws, we use $cds_1 =_{cds,Main} cds_2$ to denote the equivalence
of sets of class declarations $cds_1$ and $cds_2$, where $cds$ is a
context of class declarations for $cds_1$ and $cds_2$. We need to stress
that this definition takes into account only sequential programs. We
write `$\rightarrow$' to indicate the condition that need to be
satisfied to apply a law from left to right. Likewise, we use
`$\leftarrow$' to indicate what has to be satisfied when applying the
law from right to left. We use~`$\leftrightarrow$ to indicate conditions that must hold in both directions.

The first law we present
(\refJmlLaw{\ref{law:move_invariant_superclass}}) allows us to  move an
invariant $\psi_2$ from a subclass $C$ to its superclass $B$. The
invariant we want to move only refers to instances of $C$ as we require
the invariant to be applicable only to instances of class $C$. To apply
this law in any direction, we require that calls to $\Super$ do not
occur in $\psi_2$, since after law application (in both directions)
these calls may refer different elements. To apply this law from left to
right, model fields cannot appear in $\psi_2$ and occurrences of $\This$
must be cast otherwise the elements they refer may not be visible.

Concerning the soundness of this law, we take in account the inheritance
of specifications in JML~\cite{DBLP:conf/icfem/Leavens06}, in which
inherited invariants are conjoined with locally added invariants. On the
left-hand side, the invariant $\psi_2$, which is present in class $C$,
is inherited by the subclasses of $C$ and holds for all subclasses. On
the right-hand side of the law, the invariant $\psi_2$ is inherited by
all subclasses of $B$ besides those that are not subclasses of $C$. For
those classes that are subclasses of $B$, but not subclasses of $C$, the
invariant holds because for objects of these classes the antecedent
$\instanceof\ C$ fails and the whole implication is true, not changing
the meaning of any original local invariant that inherits~$\psi_2$.

By using
\refJmlLaw{\ref{law:move_attribute_to_superclass_cons_contracts}}, we
can move an attribute to a superclass if it is not already declared in
the superclass and if it does not cause name conflicts. The application
of \refJmlLaw{\ref{law:move_attribute_to_superclass_cons_contracts}},
from right to left, allows us to move an attribute downward. In this
case, we prevent access to the attribute by the expression $\This$, and
we allow only accesses to $a$ by $C$ or subclasses of $C$, including
accesses that appear in specifications.


\begin{figure}[t]
\begin{LawJml}{move reference type attribute to superclass}
\label{law:move_attribute_to_superclass_cons_contracts}%
\begin{framedCode} 
\cc{\class B \extends A \{}
  \cc{ads cnts mds}     
\cc{\classEnd}
\cc{\class C \extends B \{}
  \cc{\specStart \nullable \specEnd T a;}
  \cc{ads' cnts' mds'}     
\cc{\classEnd}
\end{framedCode}%
\codeIsEqualToInContext{cds}{Main}
\begin{framedCode} 
\cc{\class B \extends A \{}
  \cc{\specStart \nullable \specEnd T a;}
  \cc{ads cnts mds}   
\cc{\classEnd}
\cc{\class C \extends B \{}
  \cc{ads' cnts' mds'}   
\cc{\classEnd}
\end{framedCode}%

\begin{description}
\item[\textbf{provided} ]
\providedJml
\item [($\leftarrow$)] $D.a$, for any $D$ \subclassOp $B$ and $D$ \notSubclassOp $C$ does not occur inside specifications of $cds$, $Main$, $cnts$, $cnts$\textit{'}, $mds$ nor $mds$\texttt{'}. 
\providedJava
\item[($\leftrightarrow$)] $T$ \underline{is not} a primitive type.
\item [($\rightarrow$)] (1) $a$ is not declared in $ads$;
(2) The attribute name a is not declared by the subclasses of B in $cds$.
\item [($\leftarrow$)] $D.a$, for any $D$ \subclassOp $B$ e $D$ \notSubclassOp $C$ does not occur in $cds$, $Main$, $cnts$, $cnts$\textit{'}, $mds$ nor $mds$\texttt{'}. 
\end{description}
\end{LawJml}

\end{figure}


In  \refJmlLaw{\ref{law:move_attribute_to_superclass_cons_contracts}},
we consider  only attributes whose type is a reference type. There is
another law  for moving an attribute of primitive type. The reason for
having two disctinct laws for dealing with attributes of primitive and
reference types comes from the $\nullable$ keyword in
\refJmlLaw{\ref{law:move_attribute_to_superclass_cons_contracts}}. In
JML, any declaration (except for local variables) whose type is a
reference type is implicitly declared to be non-null, except when one
adorns the declaration with a $\nullable$ annotation~\cite{JmlRefMan}.
Thus, by default, JML always checks if a not nullable attribute is null
in all visible states of the class that declares it. When we move an
attribute to a superclass, this is not aware about the newly moved
attribute and, therefore, this action can cause a undesirable behavior.
In fact, if one instantiates the superclass, JML will raise an invariant
exception reporting that the new attribute is null. To avoid this, we
force attribute nullability to move it up. Then, if one wants to move a
non-null a attribute, one needs to introduce $\nullable$ annotation
before moving it. An attribute can become nullable applying a law named
\textit{make attribute nullable}, not presented here. Remember that in
Java only reference types can be null.

\refJmlLaw{\ref{law:move_redef_meth_to_superclass_cons_contract}}
allows us to move a redefined method from a class to its superclass. The
proviso concerning $\Super$ is needed because its semantics may be
affected when we move it from a subclass to a superclass, or vice-versa.
We can only move the specification of a method if it does not refer to
model fields of the class in which the method is originally declared.
Furthermore, $\This$ expressions may occur in the target method
specifications only if they are cast. In fact, as in the law the method
has default visibility, only non-private elements can be referenced in
its pre- and postconditions. This is similar to Java: the $\This$
expression may appear in $mbody\textit{'}$ if they have a cast and they
mention only non-private attributes or methods of class $C$. The
right-hand side of
\refJmlLaw{\ref{law:move_redef_meth_to_superclass_cons_contract}}
introduces type tests in each one of the  specifications. In
this way we assure that the original pre- and postconditions of the
redefined method of $C$ will only be applied to  callers that are
instances of $C$ or instances of any of $C$'s subclasses.


\begin{LawJml}{move redefined method to superclass: overriden method with non-default specification case}
\label{law:move_redef_meth_to_superclass_cons_contract}%
\begin{framedCode} 
\cc{\class B \extends A \{}
  \cc{ads cnts mds} 
  \cc{//@ \requires} \subescrito{\psi}{1}\cc{;}
  \cc{//@ \ensures} \subescrito{\psi}{2}\cc{;}
  \cc{rt m(pds) \{ mbody \}} 
\cc{\classEnd}
\cc{\class C \extends B \{}
  \cc{ads' cnts' mds'}  
  \cc{//@ \also}
  \cc{//@ \requires} \subescrito{\psi'}{1}\cc{;}
  \cc{//@ \ensures} \subescrito{\psi'}{2}\cc{;}
  \cc{rt m(pds) \{ mbody' \}}
\cc{\classEnd}
\end{framedCode}%
     \codeIsEqualToInContext{cds}{Main}
\begin{framedCode}
\cc{\class B \extends A \{}
  \cc{ads cnts mds}  
  \cc{//@ \requires} \subescrito{\theta}{1} && \subescrito{\psi}{1}\cc{;}
  \cc{//@ \ensures} \subescrito{\theta}{1} && \subescrito{\psi}{2}\cc{;}
  \cc{//@ \also}
  \cc{//@ \requires} \subescrito{\theta}{2} && \subescrito{\psi'}{1}\cc{;}
  \cc{//@ \ensures} \subescrito{\theta}{2} && \subescrito{\psi'}{2}\cc{;}
  \cc{//@ \also}
  \cc{//@ \requires} \subescrito{\theta}{2} && \subescrito{\psi}{1}\cc{;}
  \cc{//@ \ensures} \subescrito{\theta}{2} && \subescrito{\psi}{2}\cc{;}
  \cc{rt m(pds) \{}
     \cc{\If (!(\This \instanceof C))}
     \cc{\{ mbody \} \Else \{ mbody' \}}
  \cc{\}}
\cc{\classEnd}
\cc{\class C \extends B \{}
  \cc{ads' cnts' mds'} 
\cc{\classEnd}
\end{framedCode}%

\begin{description}
\item[\textbf{where}]
   \item[] $\theta_1 \equiv\  !(\This\ \instanceof \ \cc{C})$  and  $\theta_2 \equiv\  \This\ \instanceof \ \cc{C}$\\
   
\item[\textbf{provided}]

\providedJml
\item [($\leftrightarrow$)] $\Super$ does not appear in $\psi'_1$ nor in $\psi'_2$.

\item [($\rightarrow$)] Both $\psi_1$ and $\psi_2$ do not contain occurrences of model fields declared in $C$, nor uncast occurrences of $\This$. 

\providedJava
\item [($\leftrightarrow$)] 
(1) $\Super$ and private attributes do not appear in $mbody'$; (2) $\Super . m$ does
not appear in $mds\textit{'}$

\item [($\rightarrow$)] $mbody'$ does not contain uncast occurrences of $\This$ nor expressions of the form $((C) \This).a$ and of the form $((C) \This).m(e)$ for any attribute $a$ nor method $m$, in $ads'$ and $mds'$, respectively, with private visibility.

\item [($\leftarrow$)] $m(pds)$ is not declared in $mds'$.

\end{description}

\end{LawJml}

\section{Soundness}
\label{section:soundness}

The proofs we present here are only concerned with the JML parts of the
laws. Proving the Java part is difficult due to aliasing, which can lead
to representation exposure problems, for instance.
In JML, specifications present in a class are inherited by its
subclasses, provided they are not private. This leads us to two
concepts: join of specifications and specification inheritance.

\subsection{Join of specifications}

In a program written in Java and annotated with JML, classes inherit not
only attributes and methods from superclasses, they also inherit
specifications of invariants, methods, history constraints, and
initialisation
predicates~\cite{DBLP:conf/icfem/Leavens06,Leavens-Naumann06}.
Concerning methods, a method specification may consist of several
specifications cases, which are introduced by the use of clauses such as
\lstinline!requires!, \lstinline!assignable!,
\lstinline!ensures!~\cite{JmlRefMan}. Each specification case has a
precondition (the default predicate is \textit{true}) that states when
the corresponding specification case applies to a call. The keyword
\lstinline{also} joins specifications cases. When a precondition of a
specification case holds, the corresponding postcondition must hold
also. 
defined earlier in this section. The definitions we present here are
taken from~\cite{Leavens-Naumann06}. The notation $T \triangleright
(pre, post)$ is related to a specification case of an instance method
that type checks when its receiver (\lstinline!this!) has static type
$T$. It also type checks in contexts where \lstinline!this! has some
subtype of $T$. In what follows, we introduce the definition of joint 
JML method specifications~\cite{Leavens-Naumann06}.

\begin{mydefinition} \textbf{(Join of JML method specifications)} Let $T' \triangleright (pre', post')$ and
$T \triangleright (pre, post)$ be specifications of an instance method $m$. Let $U$ be a subtype
of both $T'$ and $T$. Then the join of $(pre', post')$ and $(pre, post)$ for $U$, written
$(pre', post')  \sqcup^U (pre, post)$, is the specification $U \triangleright (p,q)$ with precondition $p$:
\begin{center}
$pre\ ||\ pre'$
\end{center}

\noindent and postcondition $q$:
\begin{center}
($\backslash$\lstinline!old!(pre') ==$>$ post') $\&\&$ ($\backslash$\lstinline!old!(pre) ==$>$ post) 
\end{center}
\hfill $\square$
\label{def:joinSpec}
\end{mydefinition}

In Definition~\ref{def:joinSpec}, the precondition of joint method specifications
is their disjunction. The postcondition of the join is a conjunction of implications
(written ==$>$ in JML's notation), stating that when a precondition holds (in the
pre-state), the corresponding postcondition must hold.

\begin{figure}[t]\centering
\begin{tabular}{l}
\quad $ext\_inv^B_{LHS}$  \\
=  [by  Definition~\ref{def:extSpecif} ] \\
\quad $\bigwedge \{  added\_inv^U \mid U \in supers(B_{LHS}) \}$ \\
= [by set theory] \\
\quad $\bigwedge \{  added\_inv^U \mid U \in ((supers(B_{LHS}) \setminus supers(A)) \cup supers(A) \} $\\
= [by definition of conjunction]\\
\quad $(\bigwedge \{  added\_inv^U \mid U \in ((supers(B_{LHS}) \setminus supers(A)) \}) 
  \wedge (\bigwedge \{  added\_inv^W \mid W \in  supers(A) \})$  \\
=  [by definition of added invariant in $B_{LHS}$] \\
\quad $ \psi_1 \land (\bigwedge \{  added\_inv^W \mid W \in  supers(A) \})$\\
= [by Propositional Logic]\\
\quad $ \psi_1 \wedge true \wedge (\bigwedge \{  added\_inv^W \mid W \in  supers(A) \})$  \\
=  [by Propositional Logic]\\
\quad $ \psi_1 \wedge (false \Rightarrow \psi_{inv}) \wedge (\bigwedge \{  added\_inv^W \mid W \in  supers(A) \}) $ \\
= [by  type test for object of type $B$] \\
\quad $ \psi_1 \wedge (this\ instanceof\ C \Rightarrow \psi_{inv}) \wedge (\bigwedge \{  added\_inv^W \mid W \in  supers(A) \}) $ \\
=  [by definition of added invariant in $B_{RHS}$]\\
\quad $ (\bigwedge \{  added\_inv^U \mid U \in ((supers(B_{RHS}) \setminus supers(A)) \})  \wedge (\bigwedge \{  added\_inv^W \mid W \in  
supers(A) \}) $ \\
=  [by definition of conjunction] \\
\quad $ \bigwedge \{  added\_inv^U \mid U \in ((supers(B_{RHS}) \setminus supers(A)) \cup supers(A) \} $ \\
=  [by  set theory] \\
\quad $   \bigwedge \{  added\_inv^U \mid U \in supers(B_{LHS}) \} $\\
=  [by Definition~\ref{def:extSpecif} ]\\
\quad $ext\_inv^B_{RHS} $
\end{tabular}

\caption{\label{proof:Law1_typeB}Proof of \refJmlLaw{\ref{law:move_invariant_superclass}} -  case of object of exact type $B$}
\end{figure}

\subsection{Specification Inheritance}

Subtypes in JML inherit specifications, besides attributes and methods. 
First, we introduce some notation for type specification. For a type $T$, the invariant
predicate declared in the specification of $T$ (without inheritance) is denoted by
$added\_inv^T$. For a method $m$ declared in a type $T$, 
the notation $added\_spec_m^T = (added\_pre_m^T, added\_post_m^T)$ is the
join of the specification cases in type $T$ for $m$. If $m$ is declared in $T$ with
no specification and is not overriding any method, then $added\_spec_m^T = (true, true)$,
which is the default specification in JML. We use $supers(T)$ to denote the set of
all supertypes of $T$ (including $T$)  and $methods(\mathcal{T})$ to denote the set of all
instance method names declared in the specifications of the types in a set $\mathcal{T}$.

\begin{mydefinition}\textbf{(Extended specification)} Suppose T has supertypes $supers(T)$,
which includes $T$ itself. Then the extended specification of $T$ is a specification such that:

\noindent \textbf{methods:} for all methods $m\ \in methods(supers(T))$, the extended
specification of $m$ is the join of all added specifications for $m$ in $T$ and all
its proper supertypes
\begin{center}
$ext\_spec_m^T = \bigsqcup^T \{added\_spec_m^U \mid U \in supers(T) \}$
\end{center}

\noindent \textbf{invariant:} the extended invariant of $T$ is the conjunction of all
added invariants in $T$ and its proper supertypes
\begin{center}
$ext\_int^T = \bigwedge^T \{added\_inv^U \mid U \in supers(T) \}$
\end{center}
\hfill $\square$
\label{def:extSpecif}
\end{mydefinition}

The definitions we present here were introduced in~\cite{Leavens-Naumann06}
and are the ones we use in this paper.
      
\begin{figure}[t]\centering
\begin{tabular}{l}
\begin{tabular}{l}
$\quad ext\_spec_m^{B_{LHS}}$  \\
=  [by  Definition~\ref{def:extSpecif} ] \\
$\quad \bigsqcup^B_{LHS} \{added\_spec_m^U \mid U \in supers(B) \} $\\
= [by set theory]\\
$\quad \bigsqcup^B_{LHS} \{added\_spec_m^U \mid U \in (supers(B) \setminus supers(A)) \cup supers(A) \} $\\
= [by definition of join with respect to $B$]\\
$\quad (\bigsqcup^B_{LHS} \{added\_spec_m^U \mid U \in (supers(B) \setminus supers(A)) \} ) \sqcup^B (\bigsqcup^A  \{added\_spec_m^W \mid W \in supers(A) \})$\\
= [by definition of join of specification cases for $B_{LHS}$]\\
$\quad (\psi_1, \backslash old(\psi_1) \Rightarrow \psi_2 ) \sqcup^B (\bigsqcup^A  \{added\_spec_m^W \mid W \in supers(A) \})$\\
= [by Propositional Logic]\\
$\quad ((\psi_1 \wedge true ), \backslash old(\psi_1 \wedge true ) \Rightarrow \psi_2 ) 
\sqcup^B (\bigsqcup^A  \{added\_spec_m^W \mid W \in supers(A) \})$\\
= [by type test for object of type $B$]\\
$\quad ((\psi_1 \wedge \neg(this\ instanceof\ C)), \backslash old(\psi_1 \wedge \neg(this\ instanceof\ C) ) \Rightarrow \psi_2 ) $\\
$\quad \sqcup^B (\bigsqcup^A  \{added\_spec_m^W \mid W \in supers(A) \})$\\
= [by Propositional Logic]\\
$\quad ( (\psi_1 \wedge \neg(this\ instanceof\ C)) \vee false \vee false,  \backslash old(\psi_1 \wedge \neg(this\ instanceof\ C) ) \Rightarrow \psi_2 ) \wedge true \wedge true) $\\
$\quad \sqcup^B (\bigsqcup^A  \{added\_spec_m^W \mid W \in supers(A) \})$\\
= [by type test for object of type $B$ and Propositional Logic]\\
$\quad ( (\psi_1 \wedge \neg(this\ instanceof\ C)) \vee ((this\ instanceof\ C) \wedge \psi_1) \vee ((this\ instanceof\ C) \wedge \psi_1' ) , $\\
$\quad \quad (\backslash old(\psi_1 \wedge \neg(this\ instanceof\ C) ) \Rightarrow \psi_2 ) 
											 \wedge (\backslash old((this\ instanceof\ C) \wedge \psi_1 )\Rightarrow \psi_2 )) $\\
                                           $ \quad \quad \wedge (\backslash old((this\ instanceof\ C) \wedge \psi_1' )\Rightarrow \psi_2' )) 
 \sqcup^B (\bigsqcup^A  \{added\_spec_m^W \mid W \in supers(A) \})$\\
= [by definition of join of specification cases for $B_{RHS}$]\\
$\quad (\bigsqcup^B_{RHS} \{added\_spec_m^U \mid U \in (supers(B) \setminus supers(A)) \} )
\sqcup^B (\bigsqcup^A  \{added\_spec_m^W \mid W \in supers(A) \})$\\
= [by definition of join with respect to $B$]\\
$\quad \bigsqcup^B_{RHS} \{added\_spec_m^U \mid U \in (supers(B) \setminus supers(A)) \cup supers(A) \} $\\
= [by set theory]\\
$\quad \bigsqcup^B_{RHS} \{added\_spec_m^U \mid U \in supers(B) \} $\\
=  [by  Definition~\ref{def:extSpecif} ] \\
$\quad ext\_spec_m^{B_{RHS}}$  \\
\end{tabular}
\end{tabular}
\caption{\label{proof:Law3_typeB} Proof of \refJmlLaw{\ref{law:move_redef_meth_to_superclass_cons_contract}} - case of object of exact type $B$}
\end{figure}

\subsection{Proofs}

Here we present proofs for \textbf{Laws 1}  and \textbf{3}. Both proofs involve
dealing with cases associated to the types of objects related to the
classes that are emphasized in the laws. We present the proof for just
one case of these laws. The conditions of the laws guarantee that both
programs that appear in the laws are well-typed.  Concerning
\refJmlLaw{\ref{law:move_attribute_to_superclass_cons_contracts}}, it is
a law for attributes in which specification  inheritance is not taken
into consideration.

In Figure~\ref{proof:Law1_typeB}, we present the proof for the case of
\refJmlLaw{\ref{law:move_invariant_superclass}} in which the we consider
an object of exact type $B$. Notice that in
\refJmlLaw{\ref{law:move_invariant_superclass}}, on the left-hand side,
an object of exact type $B$ has to establish the (added) invariant
$\psi_1$. The added invariant is given by $\psi_1 \land (this\
instanceof\ C \Rightarrow \psi_{inv})$, on the right-hand side. For an
object of type $B$, the type test is false and the whole implication
results  true. The whole effect is the same of the invariant of class
$B$ on the left-hand side.

The proof for the case of
\refJmlLaw{\ref{law:move_redef_meth_to_superclass_cons_contract}} in
which we consider an object of exact type $B$ is presented in
Figure~\ref{proof:Law3_typeB}. On the left-hand side of this law, the
specification case for method $m$ in class $B$ has precondition $\psi_1$
and postcondition $\psi_2$. On the right-hand side, we enrich this
specification case with type tests involving the class name $C$, but
with no impacts for objects with distinct types from $C$. The other
specification case for method $m$ on the right-hand side also involves a
type test, having no effects for classes other than class $C$.

\section{Application}
\label{section:application}

\begin{figure}[t]
\begin{lstlisting}[frame=single]
	public class EvenIntegerData {
		//@ private invariant value.intValue() % 2 == 0; (*@\label{jml_example_even_integer:inv1}@*)
		//@ private invariant value.intValue() > -1; (*@\label{jml_example_even_integer:inv2}@*)
		private Integer value;
		public EvenIntegerData() { value = new Integer(0); }
		/*@ requires newValue != null; 
		  @ requires newValue.intValue() % 2 == 0 && newValue.intValue() > -1;
		  @ ensures getValue().intValue() == newValue.intValue(); @*/
		public void registerValue(Integer newValue) { /* ... */ }
    //@ ensures \result != null;
		public /*@ pure @*/ Integer getValue() { /* ... */ }		
		/*@ requires getValue() != null;
		  @ ensures !(\result).equals(""); @*/	
		public String format() { /* ... */ }
	}
\end{lstlisting}
\caption{\label{class:EvenInteger}Class  \lstinline!EvenIntegerData!}
\end{figure}

In this section, we present an example composed of excerpts of Java classes annotated 
with JML, as it is refactored by means of successive application of programming laws. 
Classes  \lstinline!PositiveIntegerData!  (Figure~\ref{class:PositiveInteger}) and \lstinline!EvenIntegerData! (Figure~\ref{class:EvenInteger}) 
represent positive and even integers, respectively.  These classes are part of a software that stores and manipulates instances
of positive and even integers.

The class \lstinline!EvenIntegerData!  (Figure~\ref{class:EvenInteger}) can only hold even positive integers 
because of  the invariant written in Line~\ref{jml_example_even_integer:inv1}. And, to 
reinforce the invariant, pre-conditions of method \lstinline!registerValue! guarantee that only even and positive values are allowed.
These classes share methods that have the same functionality. Moreover, both have an attribute called \lstinline!value! to save the 
integer value of the respective data. 
To improve the design of this program (e.g. by reducing the amount of duplicated
code) and to accept new data types (e.g. odd integers), we need to change the program in a disciplined way. 
In what follows, we present a 
guideline  that leads to the same structure as obtained by applying the refactoring Extract 
Superclass~\cite{Fowler}. We do not present all derivation steps, we 
omit most of them to save space, but each step is accomplished by the application of a law.  A detailed derivation, with all
the steps and their corresponding laws,  can be found elsewhere~\cite{Freitas:MSc}.

Our starting point  is composed by the classes presented in Figures~\ref{class:PositiveInteger} and~\ref{class:EvenInteger}.
We first introduce a new class (\lstinline!IntegerData!) to be  the superclass of the existing classes,
by applying Law \lawName{class elimination}, from right to left. Then, we change the superclass of
classes \lstinline!PositiveIntegerData! and \lstinline!EvenIntegerData! to be \lstinline!IntegerData!,
by applying Law \lawName{change superclass: from Object to another class}, from left to right.
We  prepare  classes \lstinline!PositiveIntegerData! and \lstinline!EvenIntegerData!, 
for moving  attribute, invariant, and methods. 
First, we move the common attribute \lstinline!value!
to the superclass. This is accomplished by the application of a sequence of laws, beginning
with Law \lawName{change attribute visibility: from private to public}, which changes the
visibility of the attribute \lstinline!value! to public, then we change its visibility to default by the application
of another law.
In the sequence, we apply \refJmlLaw{\ref{law:move_attribute_to_superclass_cons_contracts}}
to move the attribute \lstinline!value! from class \lstinline!PositiveIntegerData! to class \lstinline!IntegerData!.
This is followed by the application of other laws to eliminate the attribute \lstinline!value! from
class \lstinline!EvenIntegerData!.
Then, we move common methods to the class \lstinline!IntegerData!. First, we apply 
Law\lawName{move original method to superclass} to move methods from class \lstinline!PositiveIntegerData!
to \lstinline!IntegerData!; then we apply  \refJmlLaw{\ref{law:move_redef_meth_to_superclass_cons_contract}}
to move redefined methods from \lstinline!EvenIntegerData! to \lstinline!IntegerData!.
Another law allows us to simplify conditional commands.
After moving the attribute \lstinline!value! and methods to the superclass \lstinline!IntegerData!, we change the invariant
of classes \lstinline!PositiveIntegerData! and \lstinline!EvenIntegerData! to be in the format required by \refJmlLaw{\ref{law:move_invariant_superclass}}.
We apply \refJmlLaw{\ref{law:move_invariant_superclass}}  twice, then we simplify the invariant in class
\lstinline!IntegerData! and change its visibility.

In Figure~\ref{jml_example_final_version}, we  present class an excerpt of class \lstinline!IntegerData! after the application of 
the programming laws that lead to the refactoring \textit{Extract Superclass}~\cite{Fowler}.
The  final version of the \lstinline!PositiveIntegerData! and \lstinline!EvenIntegerData! has no 
\lstinline!getValue! method and does not have the common invariant (see Figure~\ref{jml_example_final_version}, 
Line~\ref{jml_example_final_version:inv}) because, now, they belong to \lstinline!IntegerData!.

\begin{figure}[t]
\begin{lstlisting}[frame=single]
	public class IntegerData {
		//@ protected invariant value.intValue() > -1; (*@\label{jml_example_final_version:inv}@*)
		protected Integer value;
			
		// @ ensures \result != null; @*/
		public /*@ pure @*/ Integer getValue() { /* ... */ }
	}
\end{lstlisting}
\caption{\label{jml_example_final_version}Class \lstinline!IntegerData!}
\end{figure}

In~\cite{Freitas:MSc}, we applied the laws  proposed  here along with
others to refactor a core module of a Manufacturing Execution System
(MES)~\cite{MES} software, which formalizes methods and procedures of
production in an integrated system and presents data in more useful and
systematic way. To control and manipulate data dynamically and in a
highly configurable way, the MES software is built on top of a Meta Data
API. We have refactored a JML-specified version of the Meta Data
API~\cite{Freitas:MSc} by applying  primitive transformations expressed
by means of our laws. We applied our laws to refactor code and to
accommodate new features. For instance, we eliminate duplicate code by
extracting a superclass that abstracts the behavior of other classes
present in the system. This is described by the \textit{Extract
Superclass} refactoring~\cite{Fowler}. Other refactorings presented by
Fowler~\cite{Fowler} (for instance, \textit{Replace Conditional with
Polymorphism} and \textit{Pull Up Method}) were also applied to the Meta
Data API by means of the proposed laws.


\section{Conclusion}
\label{section:conclusion}

Object-oriented programming laws were  proposed by Borba \textit{et
al.}~\cite{Borba04} for an object-oriented language~\cite{wps-ieee}. They proposed laws for classes and
commands; they also define a normal form for
object-oriented programs written in their language along with a reduction
strategy. They demonstrate that the set of laws is complete with
respect to this normal form. Corn\'{e}lio~\cite{CCS09,Cornelio04} proves the
laws with respect to a copy semantics~\cite{wps-ieee} and formally justifies, by using
programming laws and data refinement, refactoring practices documented
by Fowler~\cite{Fowler}. Silva, Sampaio, and Liu consider
object-oriented programming laws in a language with a reference
semantics~\cite{Leilareferencelaws}, applying such laws to code
refactoring. Duarte~\cite{Duarte08} adapts the programming laws
initially proposed in~\cite{Borba04,Cornelio04} for the Java programming and
proposes other laws for language features that are not present in the language
used in~\cite{Borba04,Cornelio04}.

In this paper, we proposed laws for object-oriented programming in the
presence of a behavioral interface specification language. 
In the laws that deal with source-doce, we treat the transformation considering the restrictions imposed by the specifications. 
These laws  are based on
programming laws  from previous work~\cite{Borba04,Cornelio04,Duarte08}
that does not consider specifications. We have considered laws that address only a subset of the JML's Level 0
constructs~\cite{JmlRefMan}, specially for lightweight specifications.
Nevertheless, our preliminary focus is to cover most of the JML
constructs that form the core notation used in the design by contract
methodology. 
We have also applied our set of laws for reducing a JML-specified Java
program to a normal form~\cite{Freitas:MSc} to address the relative
completeness of the set of laws proposed.

With respect to the order of application of programming laws in program
transformation process,  distinct orders
may lead to different results.  The conditions for application of a
programming law  usually requires the application of
other programming laws, defining that their applications are not
commutative. To obtain the desired target program, a proper law
application sequence must be established. Commonly used sequences of applications
of laws can be registered as single transformation rules.

Concerning the refactoring process, we can view a refactoring as a
target transformation that can be reached by the application of
primitive transformations expressed by means of programming laws. The
process of application of programming laws terminates when the desired
structure is reached. Corn\'{e}lio~\cite{CCS09} presents
refactorings as rules constituted by a pair of programs, similar to a law,
but the transformation described is more complex than the one of a  law. The program on
the left-hand side presents the class or classes before rule
application; the right-hand side presents the classes after rule
application. Refactoring rules 
capture complex transformations. Here we have not presented refactoring
rules as in~\cite{CCS09}, but the application of the programming
laws we presented here and in~\cite{Freitas:MSc} lead us to a result
similar to that presented in~\cite{CCS09}, that is, a refactoring
is derived by the application of programming laws.
Although at this moment our work does not provide a way to transform
programs mechanically, it offers a more reliable and
extensible alternative to address behavior-preserving transformations
like refactorings. Moreover, since some conditions present in the
laws are related to logic proofs, it is necessary to use a theorem prover as
an auxiliary tool.

Differently from laws that deal only with constructs of an
object-oriented programming language, the presence of a behavioral
interface specification language requires that we be aware of issues
related, for instance, to the visibility of specification and code
constructs, invariant preservation when introducing calls to super and
changing a parameter type to a supertype requires introducing casts in
occurrences of the parameter in specifications.

As future work, we intend to describe laws to support other
JML clauses like \keywordFont{initially}, \keywordFont{constraint},
\keywordFont{represents}, and model fields. Also, we intend to work on
proofs for the Java parts of the laws based on a reference
semantics~\cite{AN:JACM}.

\section*{Acknowledgements}

We are partially supported by the Brazilian Research Agency, CNPq, grant 484813/2007-2, and 
by the National Institute of Science and Technology for 
Software Engineering (INES\footnote{http://www.ines.org.br}), funded by CNPq and FACEPE, 
grants 573964/2008-4 and APQ-1037-1.03/08.


\bibliographystyle{eptcs} 

\begin{thebibliography}{10}
\providecommand{\bibitemstart}[1]{\bibitem{#1}}
\providecommand{\bibitemend}{}
\providecommand{\bibliographystart}{}
\providecommand{\bibliographyend}{}
\providecommand{\url}[1]{\texttt{#1}}
\providecommand{\urlprefix}{Available at }
\providecommand{\bibinfo}[2]{#2}
\bibliographystart

\bibitemstart{AN:JACM}
\bibinfo{author}{A.~Banerjee} \& \bibinfo{author}{D.~A. Naumann}
  (\bibinfo{year}{2005}): \emph{\bibinfo{title}{Ownership confinement ensures
  representation independence for \\object-oriented programs}}.
\newblock {\sl \bibinfo{journal}{Journal of the ACM}}
  \bibinfo{volume}{52}(\bibinfo{number}{6}), pp. \bibinfo{pages}{894---960}.
\bibitemend

\bibitemstart{BM97}
\bibinfo{author}{R.~Bird} \& \bibinfo{author}{O.~de~Moor}
  (\bibinfo{year}{1997}): \emph{\bibinfo{title}{Algebra of Programming}}.
\newblock \bibinfo{publisher}{Prentice Hall}.
\bibitemend

\bibitemstart{Borba04}
\bibinfo{author}{P.~Borba} et~al. (\bibinfo{year}{2004}):
  \emph{\bibinfo{title}{Algebraic reasoning for object-oriented programming}}.
\newblock {\sl \bibinfo{journal}{Science of Computer Programming}}
  \bibinfo{volume}{52}, pp. \bibinfo{pages}{53--100}.
\bibitemend

\bibitemstart{wps-ieee}
\bibinfo{author}{A.~L.~C. Cavalcanti} \& \bibinfo{author}{D.~A. Naumann}
  (\bibinfo{year}{2000}): \emph{\bibinfo{title}{A Weakest Precondition
  Semantics for Refinement of \\Object-oriented Programs}}.
\newblock {\sl \bibinfo{journal}{IEEE Transactions on Software Engineering}}
  \bibinfo{volume}{26}(\bibinfo{number}{8}), pp. \bibinfo{pages}{713--728}.
\bibitemend

\bibitemstart{Cornelio04}
\bibinfo{author}{M.~Corn{\'e}lio} (\bibinfo{year}{2004}):
  \emph{\bibinfo{title}{Refactoring as Formal Refinements}}.
\newblock \bibinfo{type}{Ph.D. thesis}, \bibinfo{school}{Universidade Federal
  de Pernambuco}.
\bibitemend

\bibitemstart{CCS09}
\bibinfo{author}{M.~Corn\'{e}lio} et~al. (\bibinfo{year}{2009}):
  \emph{\bibinfo{title}{Sound refactorings}}.
\newblock {\sl \bibinfo{journal}{Science of Computer Programming}}
  \urlprefix\url{http://dx.doi.org/10.1016/j.scico.2009.10.001}.
\bibitemend

\bibitemstart{Duarte08}
\bibinfo{author}{R.~Duarte} (\bibinfo{year}{2008}):
  \emph{\bibinfo{title}{Parallelizing Java Programs Using Transformation
  Laws}}.
\newblock \bibinfo{type}{Master's thesis}, \bibinfo{school}{Universidade
  Federal de Pernambuco (UFPE)}.
\bibitemend

\bibitemstart{Fowler}
\bibinfo{author}{M.~Fowler} (\bibinfo{year}{1999}):
  \emph{\bibinfo{title}{Refactoring: improving the design of existing code}}.
\newblock \bibinfo{publisher}{Addison-Wesley Longman Publishing Co., Inc.},
  \bibinfo{address}{Boston, MA, USA}.
\bibitemend

\bibitemstart{Freitas:MSc}
\bibinfo{author}{G.~R.~F. Freitas} (\bibinfo{year}{2009}):
  \emph{\bibinfo{title}{Refactoring Annotated Java Programs: A Rule-Based
  Approach}}.
\newblock \bibinfo{type}{Master's thesis}, \bibinfo{school}{Departamento de
  Sistemas e Computa\c{c}\~{a}o, Universidade de Pernambuco}.
\newblock \bibinfo{note}{Available
  \href{http://www.gabrielfalconieri.com.br/msc/masterThesis.pdf}{online}}.
\bibitemend

\bibitemstart{Goldstein06}
\bibinfo{author}{M.~Goldstein}, \bibinfo{author}{Y.~A. Feldman} \&
  \bibinfo{author}{S.~Tyszberowicz} (\bibinfo{year}{2006}):
  \emph{\bibinfo{title}{Refactoring with Contracts}}.
\newblock In: {\sl \bibinfo{booktitle}{AGILE '06: Proceedings of the conference
  on AGILE 2006}}. \bibinfo{publisher}{IEEE Computer Society},
  \bibinfo{address}{Washington, DC, USA}, pp. \bibinfo{pages}{53--64}.
\bibitemend

\bibitemstart{Java}
\bibinfo{author}{J.~Gosling} et~al. (\bibinfo{year}{2005}):
  \emph{\bibinfo{title}{Java Language Specification}}.
\newblock \bibinfo{publisher}{Addison-Wesley Professional},
  \bibinfo{edition}{3rd edition} edition.
\bibitemend

\bibitemstart{hoare}
\bibinfo{author}{C.~A.~R. Hoare} et~al. (\bibinfo{year}{1987}):
  \emph{\bibinfo{title}{Laws of programming}}.
\newblock {\sl \bibinfo{journal}{Commun. ACM}}
  \bibinfo{volume}{30}(\bibinfo{number}{8}), pp. \bibinfo{pages}{672--686}.
\bibitemend

\bibitemstart{DBLP:conf/icfem/Leavens06}
\bibinfo{author}{G.~T. Leavens} (\bibinfo{year}{2006}):
  \emph{\bibinfo{title}{JML's Rich, Inherited Specifications for Behavioral
  Subtypes}}.
\newblock In: \bibinfo{editor}{Z.~Liu} \& \bibinfo{editor}{J.~He}, editors:
  {\sl \bibinfo{booktitle}{ICFEM}}, {\sl \bibinfo{series}{Lecture Notes in
  Computer Science}} \bibinfo{volume}{4260}. \bibinfo{publisher}{Springer}, pp.
  \bibinfo{pages}{2--34}.
\bibitemend

\bibitemstart{LC05}
\bibinfo{author}{G.~T. Leavens} \& \bibinfo{author}{Y.~Cheon}
  (\bibinfo{year}{2005}): \emph{\bibinfo{title}{Design by Contract with
  {JML}}}.
\newblock \bibinfo{note}{Available
  \href{http://www.jmlspecs.org/jmldbc.pdf}{online}}.
\bibitemend

\bibitemstart{Leavens-Naumann06}
\bibinfo{author}{G.~T. Leavens} \& \bibinfo{author}{D.~A. Naumann}
  (\bibinfo{year}{2006}): \emph{\bibinfo{title}{Behavioral Subtyping,
  Specification Inheritance, and Modular \\ Reasoning}}.
\newblock \bibinfo{type}{Technical Report} \bibinfo{number}{06-20b},
  \bibinfo{institution}{Department of Computer Science, Iowa State University},
  \bibinfo{address}{Ames, Iowa, 50011}.
\bibitemend

\bibitemstart{JmlRefMan}
\bibinfo{author}{G.~T. Leavens} et~al. (\bibinfo{year}{2008}):
  \emph{\bibinfo{title}{JML Reference Manual}}.
\newblock \bibinfo{note}{Available
  \href{http://www.eecs.ucf.edu/~leavens/JML/jmlrefman/}{online}.}
\bibitemend

\bibitemstart{DBLP:conf/fase/MassoniGB08}
\bibinfo{author}{T.~Massoni}, \bibinfo{author}{R.~Gheyi} \&
  \bibinfo{author}{P.~Borba} (\bibinfo{year}{2008}):
  \emph{\bibinfo{title}{Formal Model-Driven Program Refactoring}}.
\newblock In: \bibinfo{editor}{J.~L. Fiadeiro} \&
  \bibinfo{editor}{P.~Inverardi}, editors: {\sl \bibinfo{booktitle}{FASE}},
  {\sl \bibinfo{series}{Lecture Notes in Computer Science}}
  \bibinfo{volume}{4961}. \bibinfo{publisher}{Springer}, pp.
  \bibinfo{pages}{362--376}.
\newblock \urlprefix\url{http://dx.doi.org/10.1007/978-3-540-78743-3_27}.
\bibitemend

\bibitemstart{Meyer92applyingdesign}
\bibinfo{author}{B.~Meyer} (\bibinfo{year}{1992}):
  \emph{\bibinfo{title}{Applying design by contract}}.
\newblock {\sl \bibinfo{journal}{IEEE Computer}} \bibinfo{volume}{25}, pp.
  \bibinfo{pages}{40--51}.
\bibitemend

\bibitemstart{Morgan:book}
\bibinfo{author}{C.~C. Morgan} (\bibinfo{year}{1994}):
  \emph{\bibinfo{title}{{Programming from Specifications}}}.
\newblock \bibinfo{publisher}{Prentice Hall}, \bibinfo{edition}{second}
  edition.
\bibitemend

\bibitemstart{Seres:thesis}
\bibinfo{author}{S.~Seres} (\bibinfo{year}{2001}): \emph{\bibinfo{title}{The
  Algebra of Logic Programming}}.
\newblock \bibinfo{type}{Ph.D. thesis}, \bibinfo{school}{Oxfor University
  Computing Laboratory}.
\bibitemend

\bibitemstart{Leilareferencelaws}
\bibinfo{author}{L.~Silva}, \bibinfo{author}{A.~Sampaio} \&
  \bibinfo{author}{Z.~Liu} (\bibinfo{year}{2008}): \emph{\bibinfo{title}{Laws
  of Object-Orientation with Reference Semantics}}.
\newblock In: {\sl \bibinfo{booktitle}{SEFM '08: Proceedings of the 2008 Sixth
  IEEE International Conference on Software Engineering and Formal Methods}}.
  \bibinfo{publisher}{IEEE Computer Society}, \bibinfo{address}{Washington, DC,
  USA}, pp. \bibinfo{pages}{217--226}.
\bibitemend

\bibitemstart{MES}
\bibinfo{author}{R.~R. Zagidullin} \& \bibinfo{author}{E.~B. Frolov}
  (\bibinfo{year}{2008}): \emph{\bibinfo{title}{Control of manufacturing
  production by means of MES systems}}.
\newblock {\sl \bibinfo{journal}{Russian Engineering Research}}
  \bibinfo{volume}{28}(\bibinfo{number}{2}), pp. \bibinfo{pages}{166--168}.
\bibitemend

\bibliographyend
\end{thebibliography}

\end{document}